\documentclass[aps,preprint,superscriptaddress]{revtex4-1}

\usepackage{graphicx}
\usepackage{amsmath}
\usepackage{amssymb}
\usepackage{slashed}

\usepackage{array} 
\newcolumntype{C}{>{$}c<{$}}

\begin{document}


\title{A coupled-channel system \\ with anomalous thresholds and unitarity}


\author{Csaba L. Korpa}
\affiliation{MTA-PTE High-Field Terahertz Research Group,\\ Ifj\'us\'ag \'utja 6,
7624 P\'ecs, Hungary, \\and \\Institute of Physics, University of P\'ecs,\\ Ifj\'us\'ag \'utja 6,
7624 P\'ecs, Hungary}
\author{Matthias F.M. Lutz}
\affiliation{GSI Helmholtzzentrum f\"ur Schwerionenforschung GmbH, \\Planckstra\ss e 1, 64291 Darmstadt, Germany}
\author{Xiao-Yu Guo}
\affiliation{Faculty of Science, Beijing University of Technology,\\
  Beijing 100124, China}
\author{Yonggoo Heo}
\affiliation{Bogoliubov Laboratory for Theoretical Physics, Joint Institute for Nuclear Research, RU-141980 Dubna, Moscow region, Russia}
\affiliation{GSI Helmholtzzentrum f\"ur Schwerionenforschung GmbH, \\Planckstra\ss e 1, 64291 Darmstadt, Germany}
\date{\today}

\begin{abstract}

We consider the isospin one-half example system, with $D \,\pi , D\,\eta, D_s \bar K, D^* \pi , D^*\eta, D^*\bar K$ coupled channels in the $J^P = 1^-$ partial wave, chosen such that 
various phenomena that come with the opening of an anomalous threshold can be illustrated in a step-wise 
procedure by a suitable variation of up, down and strange quark masses. 
We use a set of low-energy constants in the chiral Lagrangian that were adjusted to a large set of Lattice QCD results. 
The six phase shifts and inelasticity parameters are presented for various choices of the pion mass. For a pion mass of 150 MeV there are no anomalous thresholds encountered. The small change from 150 MeV to 145 MeV pion mass causes a dramatic impact of the anomalous threshold on the phase shifts. 

\end{abstract}

\pacs{12.38.-t,12.38.Cy,12.39.Fe,12.38.Gc,14.20.-c}
\keywords{Chiral extrapolation, chiral symmetry, flavor $SU(3)$, charmed mesons, Lattice QCD}

\maketitle



\section{Introduction}
\label{sec:1} 

A system that is subject to strong interactions is typically characterized by a set of not-necessarily-well-separated scales, which reflect its dynamical features. This is contrasted by the fact that the underlying fundamental theory of QCD exhibits clear scale separations at least if evaluated in perturbation theory. While with Lattice QCD technology the non-perturbative sector is more and more accessible, the unravelling of the various sales is quite a challenging enterprise. Here effective field theory (EFT) approaches can test assumptions on the active scales in given sectors of QCD. Traditionally this is explored in terms of an effective Lagrangian with a particular choice of hadronic degrees of freedom supplemented by some power counting rules that justify in the best case a perturbative evaluation of the Lagrangian. In that case a possible scale separation is part of the setup of the effective Lagrangian, i.e. no further scales are then generated dynamically. Unfortunately, the best case assumption is rarely met. Typically an EFT works best 
at kinematical points, where physical scattering processes are not possible. In case of a two-body reaction this kinematic condition is met if the three Mandelstam variables $s,\, t$ and $u$ are set to values below their respective thresholds. This is unfortunate since experimental data are collected typically outside the Mandelstam triangle, where the amplitude is characterized by possible resonance states.

Therefore it is highly desirable to extend the scope of EFT beyond its perturbative 
domain with the possibility to generate additional scales that are not manifest in the effective Lagrangian. From a mathematical point of view the solution to this problem is almost trivial, in the sense that a fundamental theorem  of analytic functions guarantees that if we know a reaction amplitude within the Mandelstam triangle, there exists a unique analytic continuation into the entire complex plane. The  
caveat is, that we would need knowledge on the amplitude at infinite precision. Clearly, such a request is  impossible to be delivered by an EFT. Or turned around, the analytic continuation from inside the Mandelstam triangle to 
outside points turns more  and more uncertain with increasing distance from the center of the triangle. 
It would appear that an EFT is unable to make any statement  on the resonance spectrum of QCD. 
However, such a conclusion is incorrect, since there is a powerful way out of this misery. Consider only such analytic continuations that are exactly consistent with the coupled-channel unitarity conditions. Within such a setup the desired analytic continuation is stabilized and the properties of resonances can be studied systematically in terms of the Generalized Potential Approach (GPA) as introduced in \cite{Gasparyan:2010xz,Gasparyan:2011yw,Danilkin:2010xd,Gasparyan:2012km}.

In this letter we present a realistic example system at the hand of which some novel features of such an extended EFT can be illustrated. Our basis is the chiral Lagrangian for the open-charm sector of QCD with three light flavors, i.e. with up, down and strange quarks, and one heavy flavor, i.e. the charm quark. There are two well-known approximate symmetries that characterize the EFT. The chiral SU(3) and the heavy-quark spin symmetry \cite{Casalbuoni:1996pg}. While the chiral symmetry explains the pseudo-Goldstone nature of the pion, kaon and eta, the spin symmetry does the approximate degeneracy of the $J^P=0^-$ and $J^P=1^-$ charmed meson masses. In the pioneering work  \cite{Kolomeitsev:2003ac} it was demonstrated that with the leading order two-body interaction terms of such an EFT, if extended from the center of the Mandelstam triangle by a s- or u-channel unitarization into the resonance region, a striking set of open-charm resonance states with $J^P=0^+$ and $J^P=1^+$ quantum numbers is dynamically generated with their associated set of additional scales. In the flavor limit with degenerate  up, down and strange quark masses, a strongly bound flavor triplet and a less bound flavor sextet were predicted.  While by now there is a rich literature on such s-wave resonances (see e.g. \cite{Hofmann:2003je,Lutz:2007sk,Guo:2009ct,Albaladejo:2016lbb,Du:2017zvv,Guo:2018kno,Guo:2018gyd}), it is much less studied what happens if the two-body states sit in a relative p-wave. This is so since the assumption that the coupled-channel interaction is dominated by short range forces starts to break down in higher partial-wave contributions. Therefore unitarization methods that are able to cope with short- and long-range interactions simultaneously are required as is possible in the GPA \cite{Lutz:2022enz}.  In 
particular further work is needed in the case that anomalous thresholds arise \cite{Karplus:1958zz, Mandelstam:1960zz,Ball:Frazer:Nauenberg:1962,Greben:1976wc,Lutz:2015lca,Lutz:2018kaz}

P-wave studies probe the subleading order LEC of the chiral Lagrangian that need to be determined by some data set. Only recently the authors considered the possible dynamic generation of p-wave states with $J^P = 1^-$ as implied  by a set of LEC extracted from Lattice QCD data for the first time. Such studies are significant as they depend on a set of LEC that can be determined from Lattice QCD data at unphysically large quark masses \cite{Mohler:2011ke,Liu:2012zya,Mohler:2013rwa,Lang:2014yfa,Moir:2016srx,Cheung:2020mql,Gayer:2021xzv,Lang:2022elg}. While such a computation can be executed along the GPA introduced in \cite{Gasparyan:2010xz,Gasparyan:2011yw,Danilkin:2010xd,Gasparyan:2012km} for large quark masses, as the up and down quark masses are reduced towards the physical point conceptional challenges that require an extension of the conventional GPA along the line worked out recently in \cite{Lutz:2015lca,Lutz:2018kaz}  arise. Our p-wave target system requests 6 channel, $D\, \pi, D \,\eta,  D_s\bar K $ and $D^* \pi, D^*\eta,  D^*_s\bar K$, where depending on the choice of quark masses various reaction channels need particular attention. In our study we dial the quark masses such that the kaon masses are kept on their physical value always. A particular choice for the pion mass can then be translated back to specific values of the quarks mass $m_u = m_d$ and $m_s$. Given such a framework 
the reactions $D^* \pi \leftrightarrow  D^* \eta $ or $D^* \pi \to  D^* \pi $ develop anomalous left-hand cuts from the u-channel exchange of the $D$ meson at pion masses smaller than about 150 MeV.  In our numerical study we implemented the scheme \cite{Lutz:2015lca,Lutz:2018kaz}, for  which we show explicit results for the first time.



\section{Coupled-channel reaction amplitudes}

A coupled-channel partial-wave reaction amplitude, $T^J_{ab}(s)$, with total angular momentum $J$ is characterized by left-hand and right-hand cuts, where the right-hand cuts are implied by the s-channel coupled-channel unitarity condition. 
The generalized potential, $U^J_{ab}(s)$,  is determined by the left-hand cut contributions only. The separation is implied by the nonlinear integral equation
\begin{eqnarray}
T^J_{ab}(s)=U^J_{ab}(s)
+\sum_{c,d}\int^{\infty}_{\mu_{thr}^2}\frac{d\bar{s}}{\pi}\frac{s-\mu_M^2}{\bar{s}-\mu_M^2}\,
\frac{T^J_{ac}(\bar s)\,\rho^J_{cd}(\bar s)\,T^{J*}_{db}(\bar s)}{\bar{s}-s-i\epsilon}\,,
\label{def-non-linear}
\end{eqnarray}
with the phase-space matrix, $\rho^J_{ab} (s)$, depending on the coupled-channel indices $a$ and $b$. Turned around, for a given $U^J_{ab}(s)$ an amplitude $T^J_{ab}(s)$ that solves the nonlinear coupled-channel integral equation (\ref{def-non-linear}) satisfies the coupled-channel unitarity condition by construction.
The matching scale 
\begin{eqnarray}
&& \mu_M^2  \equiv  m_1^2 + M_1^2 +  m_1\,M_1  \,,
\label{def-matching}
\end{eqnarray}
in (\ref{def-non-linear}) specifies 
where we expect the conventional EFT approach to coincide with the non-perturbative coupled-channel approach followed here. It should be slightly below the smallest two-body threshold at $m_1+ M_1$ accessible in a sector with given isospin and strangeness. Given our approximation scheme it cannot be moved much further left, as the unitarity effects from the crossed u-channel will turn more and more important.

From the form of (\ref{def-non-linear}) it follows that the existence of a solution requires the generalized potential to be bounded asymptotically, modulo some possibe logarithmic terms.
Therefore, a direct evaluation of $U^J_{ab}(s)$ in the EFT is not achievable. Any
finite order truncation leads to an unbounded potential,
characterized by an asymptotic growth in some powers of $s$. 
However, we may split the left-hand cut contributions into 'close-by' and 'far-distant' contributions
\begin{eqnarray}
&& U(s )= U_{\rm close-by}(s) + U_{\rm far-distant}(s)   \,,
\nonumber\\
&& {\rm with } \qquad U_{\rm far-distant}(s) = \sum_{k}^{} \,c_k \,\xi^k(s)  \,,
\label{def-U-xipanded}
\end{eqnarray}
where we may expand the far-distant term by means of a conformal expansion. While the close-by contributions are obtainable by the EFT and are asymptotically bounded by construction, the far-distant 
terms are bounded also if a properly constructed conformal map is used  \cite{Gasparyan:2010xz,Danilkin:2010xd}.  They may be reconstructed unambiguously in terms of some derivatives at a chosen point, where the results of the EFT approach are reliable. For sufficiently large quark masses such a program is well-defined and documented in our recent work \cite{Lutz:2022enz}.  From a global fit to ssLattice QCD data a set of LEC 
was determined, which we rely on in this work.

\begin{figure}[t]
\vskip-0.1cm
\center{
\includegraphics[keepaspectratio,width=0.75\textwidth]{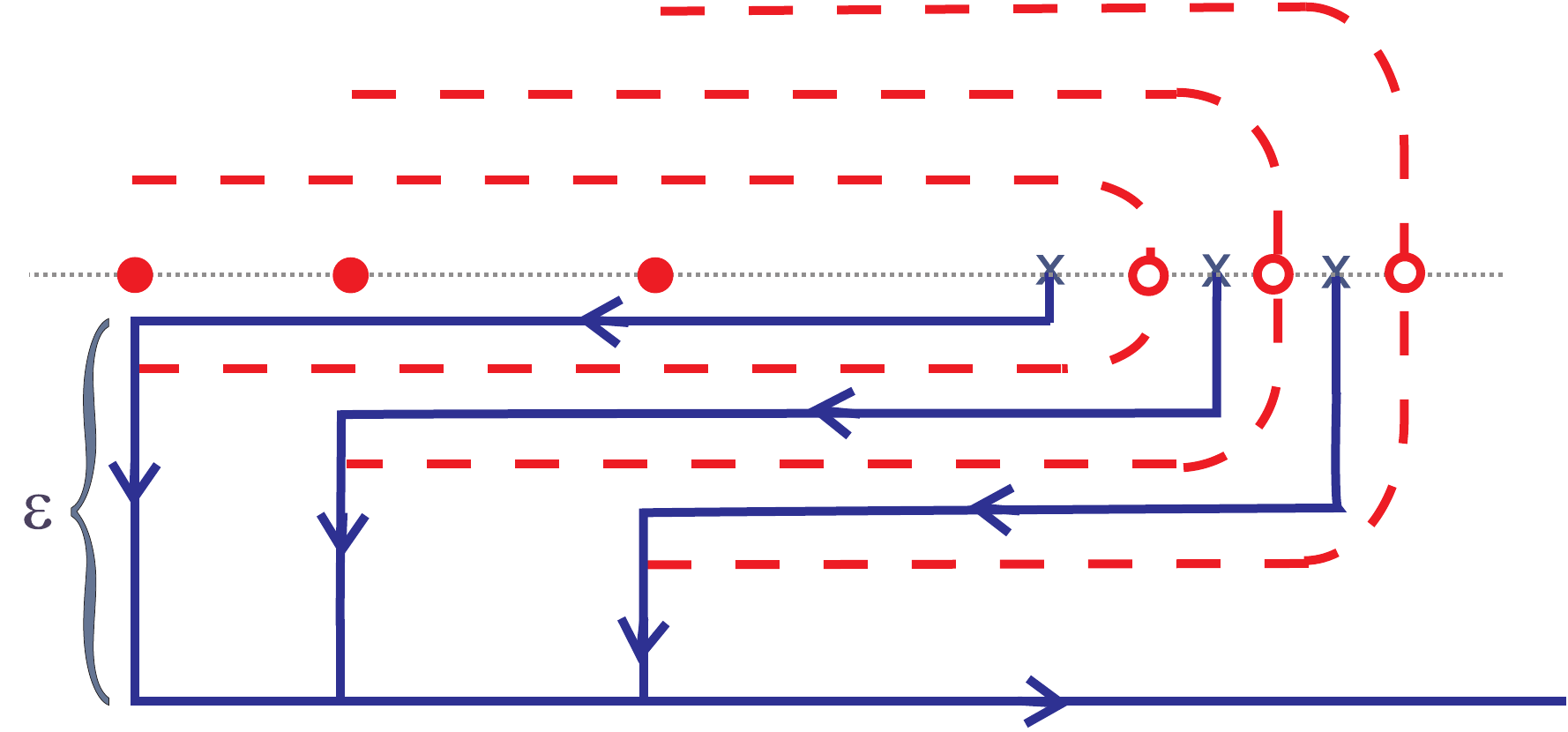} }
\vskip-0.2cm
\caption{\label{fig1} The deformed  left- and right-hand cut lines for the reaction amplitudes in (\ref{def-non-linear}). 
The crosses show the location of the normal threshold points. While the anomalous left-hand cut lines are shown with  dashed red lines, the deformed right-hand cut lines are represented by blue solid lines. The filled and open red circles indicate the location of the  anomalous threshold points and the return points   respectively as explained in \cite{Lutz:2018kaz}. }
\end{figure}

Matters turn more complicated in the presence of anomalous thresholds \cite{Lutz:2015lca,Lutz:2018kaz}. In the nonlinear integral equation (\ref{def-non-linear}) the integration contours have to be modified. While the normal threshold points are untouched the contour lines have to be  mended as illustrated in Fig. \ref{fig1}. For a normal system such a deformation of the contour lines does not change the reaction amplitudes evaluated above the real axis. As we lower the pion mass anomalous left-hand cut lines may develop, drawn here in dashed red. Now, a left-hand cut line may circle around a threshold point. Only with the mended solid blue lines the left-hand cut lines in dashed-red can be chosen not to cross any of the unitarity cut lines in solid blue. This implies that the use of the mended cut lines defines an analytic continuation  from the normal case to the anomalous case as is requested by our change of quark masses. While it is almost trivial to make the point, the challenge is to translate this in a stable numerical method as to actually derive the reaction amplitudes in such cases \cite{Lutz:2015lca,Lutz:2018kaz}.

A discussion of the subtle consequences of the presence of an anomalous threshold  is the main focus of the current work. We will not attempt to provide an estimate of systematic uncertainties, as our goal is the  documentation of our novel method how to tackle coupled-channel systems in the presence of anomalous thresholds. Rather than repeating the formal developments 
already established in \cite{Lutz:2015lca,Lutz:2018kaz} we present here results from our numerical implementation.

\section{Coupled-channel phase shifts }

\begin{figure}[t]
\center{
\includegraphics[keepaspectratio,width=1.0\textwidth]{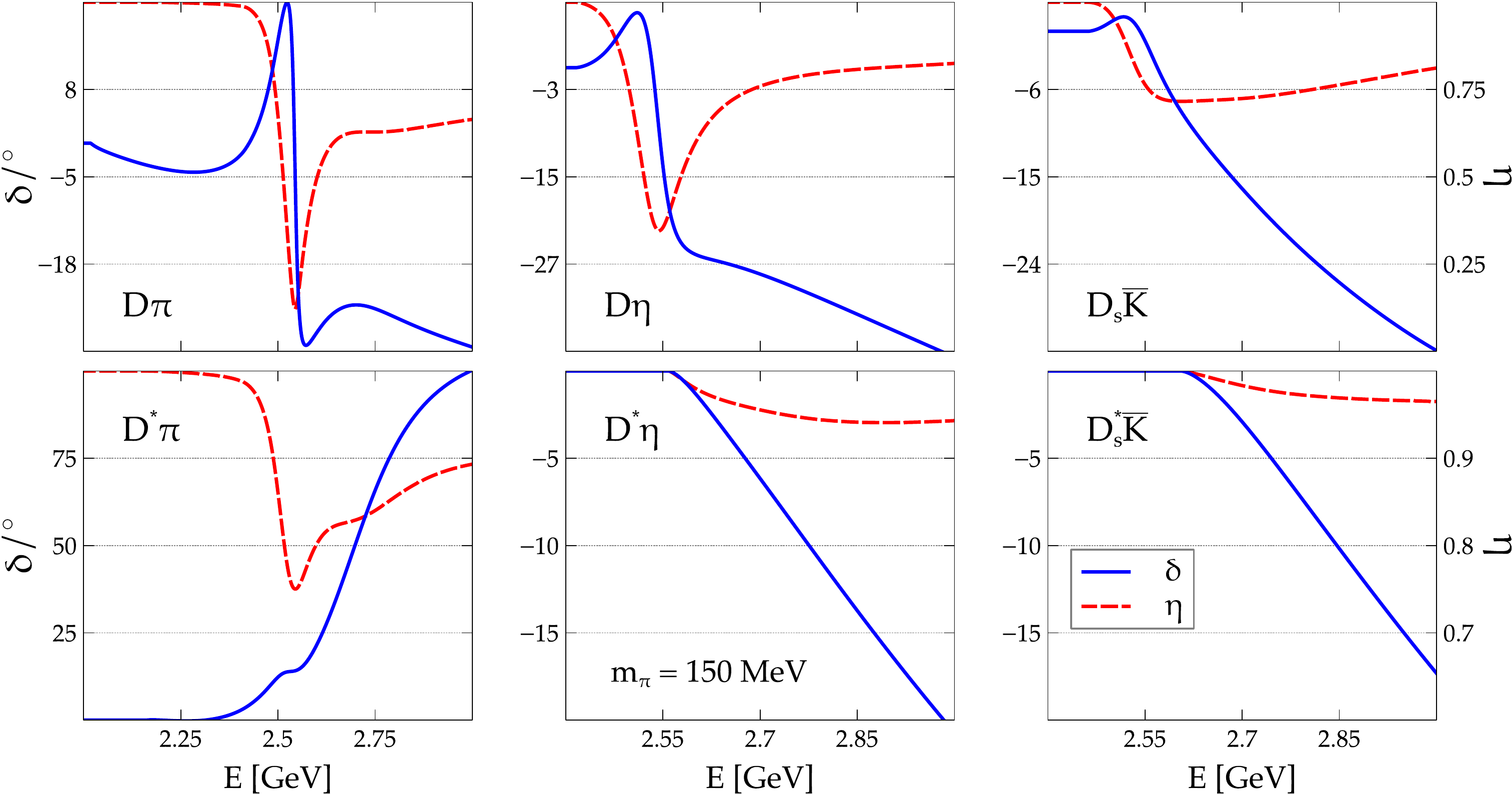}
\vskip-0.3cm
\caption{\label{fig2} P-wave phase shifts with $I=1/2$ and $J^P = 1^-$ quantum numbers.  Results are shown for the pion mass $m_\pi = 150 $ MeV, where no anomalous threshold is encountered. Phase shifts are shown in blue solid lines, inelasticities in dashed red lines.  }}
\end{figure}

Given the results of our previous work \cite{Lutz:2022enz} we can choose the pion and kaon masses such as to simplify the discussion of the anomalous threshold phenomena in the p-wave system with 
the   $D \,\pi , D\, \eta ,  D_s \bar K $ and $ D^* \pi, D^* \eta , D^*_s \bar K$ isospin one-half channels. Insisting on the physical value of the kaon mass it suffices to consider the pion mass form its physical value up to masses about 150 MeV. For the pion masses larger than 150 MeV the system is void of anomalous thresholds. For this case we show  in Fig. \ref{fig2}  all phase shifts and inelasticity parameters as implied by our global set of LEC from \cite{Lutz:2022enz}. The reaction amplitudes show a real pole on the 1st Riemann sheet at the $D^*$ mass. By construction that pole is below the $D\,\pi $ threshold, and therefore not seen in any of the phase shifts.

\begin{figure}[t] 
\center{
\includegraphics[keepaspectratio,width=0.76\textwidth]{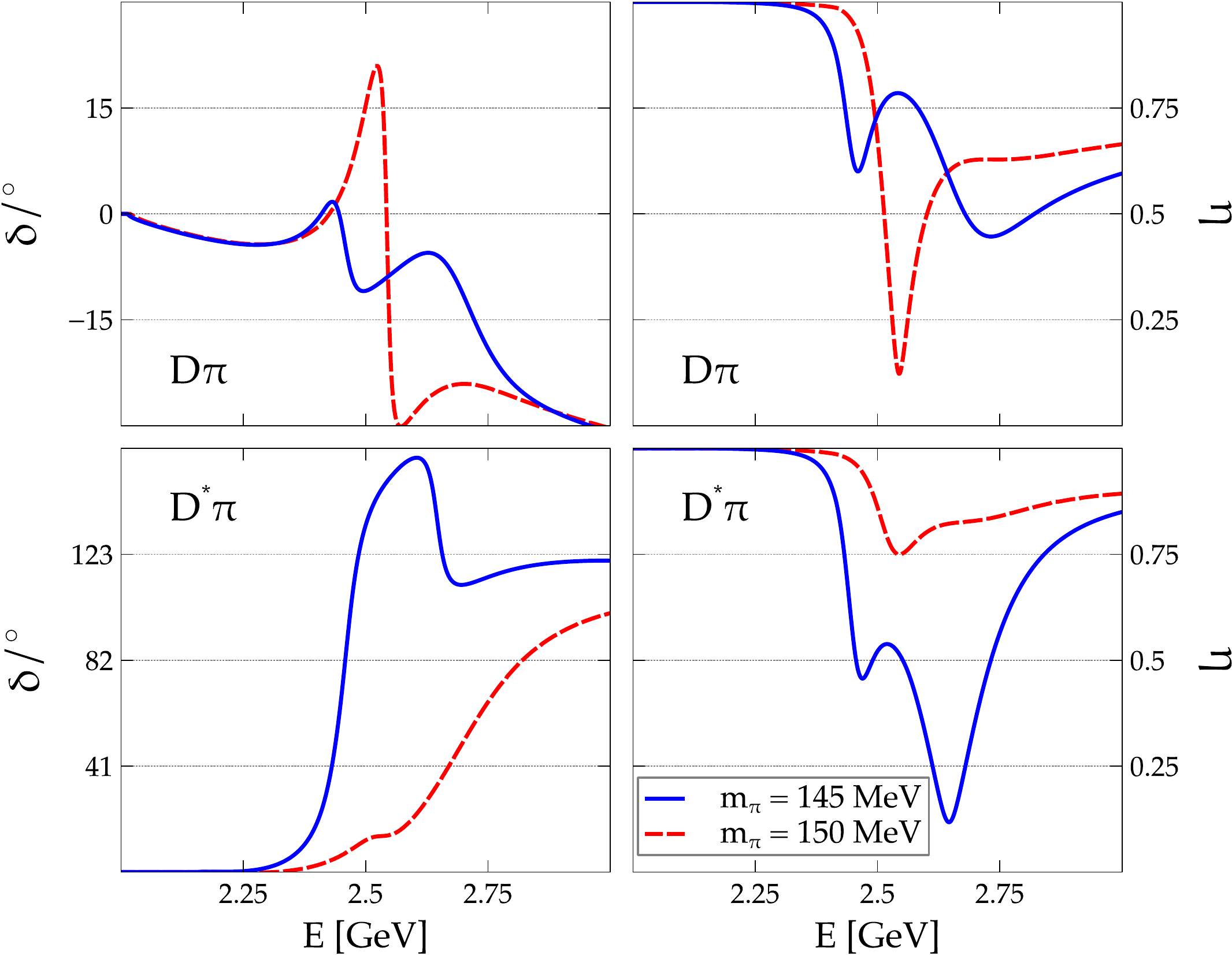}
\vskip-0.6cm
\caption{\label{fig3} P-wave phase shifts with $I=1/2$ and $J^P = 1^-$ quantum numbers. The l.h. panels show the phase shifts, the r.h.p. the inelasticity parameters for  the $D\, \pi$ and $D^* \pi$ channels. Results are shown for two choices of the pion masses, $m_\pi = 145 $ MeV and $m_\pi = 150 $ MeV. }}
\end{figure}

\begin{figure}[t] 
\center{
\includegraphics[keepaspectratio,width=0.76\textwidth]{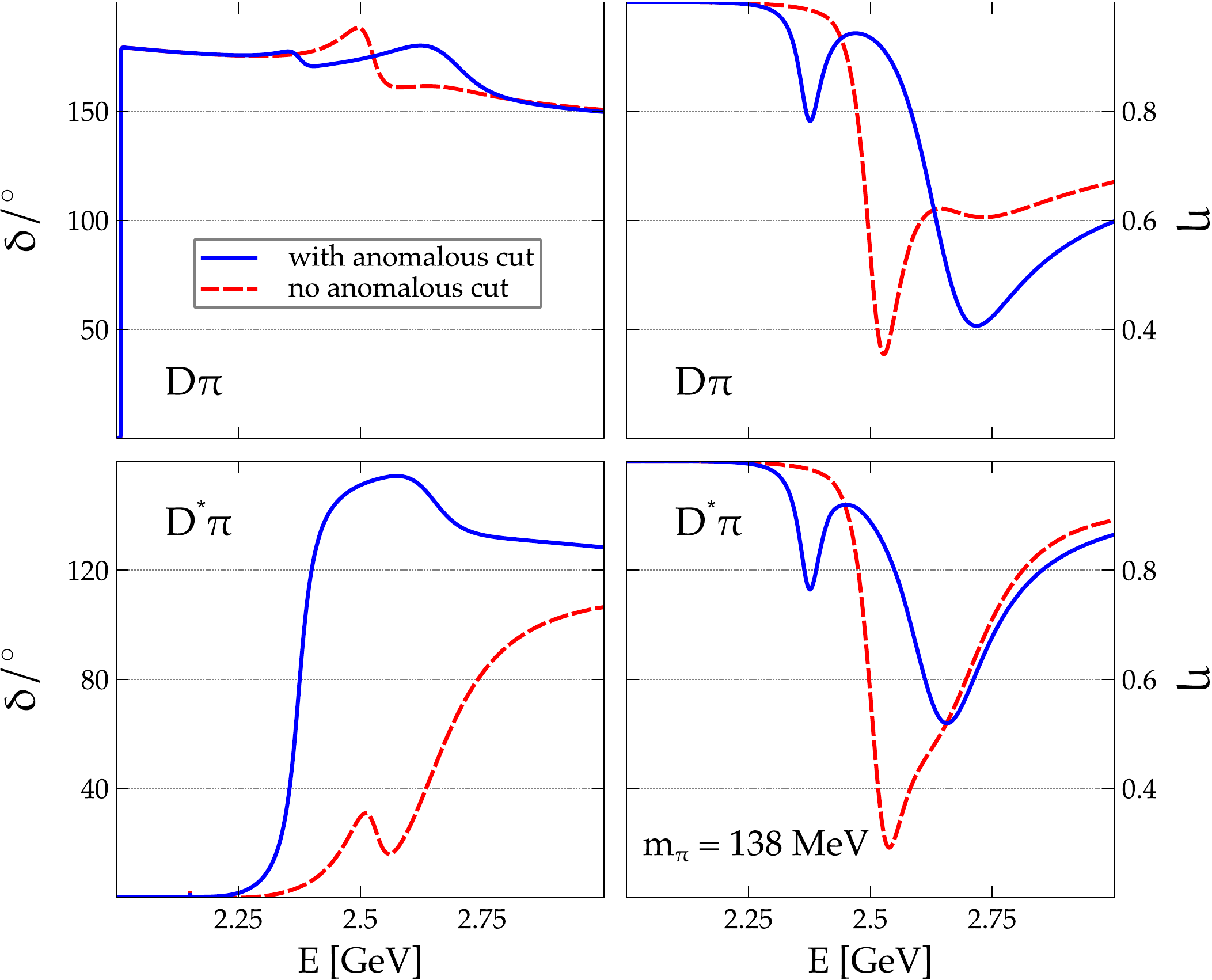}
\vskip-0.6cm
\caption{\label{fig4} P-wave phase shifts with $I=1/2$ and $J^P = 1^-$ quantum numbers. The l.h. panels show the phase shifts, the r.h.p. the inelasticity parameters for  the $D\, \pi$ and $D^* \pi$ channels. Results are shown for the physical pion mass}}
\end{figure} 

As a first novel and striking result we compare the normal system at pion mass 150 MeV with the 
anomalous system at pion mass of 145 MeV in Fig. \ref{fig3}. Despite the minor change in the pion mass, for the phase shift and and inelastiticty parameters  in  the 
$ D\, \pi$ and $D^* \pi $ channels, the opening of the anomalous threshold in the $D^* \pi\leftrightarrow D^* \eta$ channel from the u-channel exchange of the $D$ meson has a large impact on the form of the reaction amplitudes. We do not show here phase shifts and inelasticity parameters in channels where 
such a small change in the pion mass is of minor importance.  
It was explicitly checked that our numerical results are compatible with the coupled-channel unitarity condition for both cases, in particular for the case encountering the anomalous threshold. To the best knowledge of the authors such an effect has not been documented in the literature before. We emphasize that for both choices of the pion mass the $D^*$ meson is stable against hadronic decays, and therefore the $D^* \pi$ is  a clean two-body state.

In the final Fig. \ref{fig4} we show our results for the physical pion mass. In this case the left-hand cut in the  $D^* \pi \to D^* \pi $ channel  from the u-channel exchange of the $D$ meson is characterized by branch points above the $D^* \pi$ threshold. 
We illustrate the role of the anomalous cut in the $D^* \pi \leftrightarrow  D^* \eta$ channel. The solid blue lines show our result in the presence of such a cut, the dashed red lines in the absence of it. 
Switching this contribution off causes a dramatic change in the $D \,\pi$ and $D^* \pi$ phase shifts. We note that our final result for both phase shifts is quite close to the results at $m_ \pi = 145$ MeV, where the 
left-hand cut is normal  in  the  $D^* \pi \to D^* \pi $ channel with branch points below the $D^* \pi$ threshold.

\section{Summary}

In this letter we illustrated the impact of an anomalous threshold in a specific coupled-channel system. 
Such systems occur frequently in hadron physics particularly in Lattice QCD studies, where the light quark masses can be changed off their physical values. Sustainability and economic efficiency suggest to start such computations at unphysically large quark masses and then chirally extrapolate down the results to the physical point. We have shown that such extrapolations are possible, however, they require more sophisticated technology, in particular, if the system goes through anomalous regions. Small changes in the pion mass may cause dramatic effects in the phase shift, and even the dynamic generation of p-wave resonances is possible and natural in this case. 

\section*{Acknowledgments}

M.F.M. Lutz acknowledges  Tobias Isken and Daniel Mohler for stimulating discussions. 

\bibliography{literature}
\bibliographystyle{apsrev4-1}
\end{document}